\documentclass[11pt]{article}
\usepackage{moriond,epsfig}

\def\be{\begin{equation}}
\def\ee{\end{equation}}
\def\bea{\begin{eqnarray}}
\def\eea{\end{eqnarray}}

\begin{document}
\vspace*{4cm}
\title{Medium properties and jet-medium interaction from STAR}

\author{Jana Bielcikova for the STAR Collaboration}

\address{Nuclear Physics Institute ASCR, Na Truhlarce 39/64, Prague, 18086, Czech Republic}

\maketitle\abstracts{We report recent STAR results on medium modification of 
2- and 3-particle correlations in heavy-ion collisions at RHIC. In particular 
properties of an additional extended correlation in pseudo-rapidity (``the ridge''), not present in p+p or d+Au collisions, are discussed. Next, the shape modification of the away-side correlation peak 
at low- and intermediate-$p_T$ is investigated to look for signals of Mach cone shock waves or \v{C}erenkov gluon emission.}

Di-hadron azimuthal correlations are commonly used to study jet-like processes 
in heavy-ion collisions at RHIC. It has been observed that both near- and away-side correlation peaks show striking differences  to p+p and d+Au measurements.  In this paper we discuss recent measurements of azimuthal ($\Delta\phi$) and pseudo-rapidity ($\Delta\eta$) correlations in the STAR experiment using 2- and 3-particle correlation techniques, identified particles and different collision systems (d+Au, Cu+Cu, Au+Au) in order to obtain more insight into the particle production mechanism at RHIC.

\begin{figure}[t!]
\begin{center}
\begin{tabular}{lr}
\includegraphics[width=6.2cm]{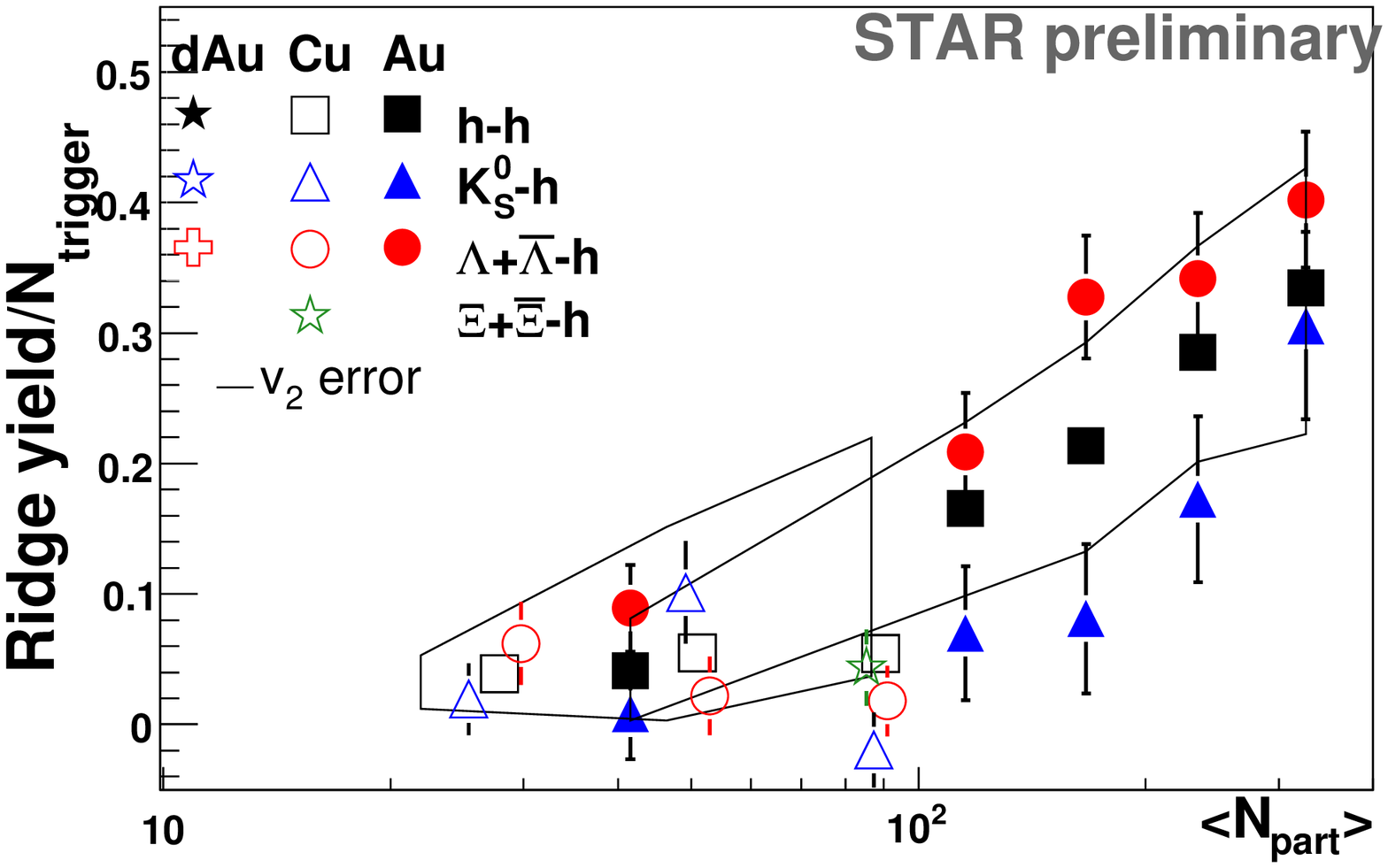}
&
\includegraphics[width=6.2cm]{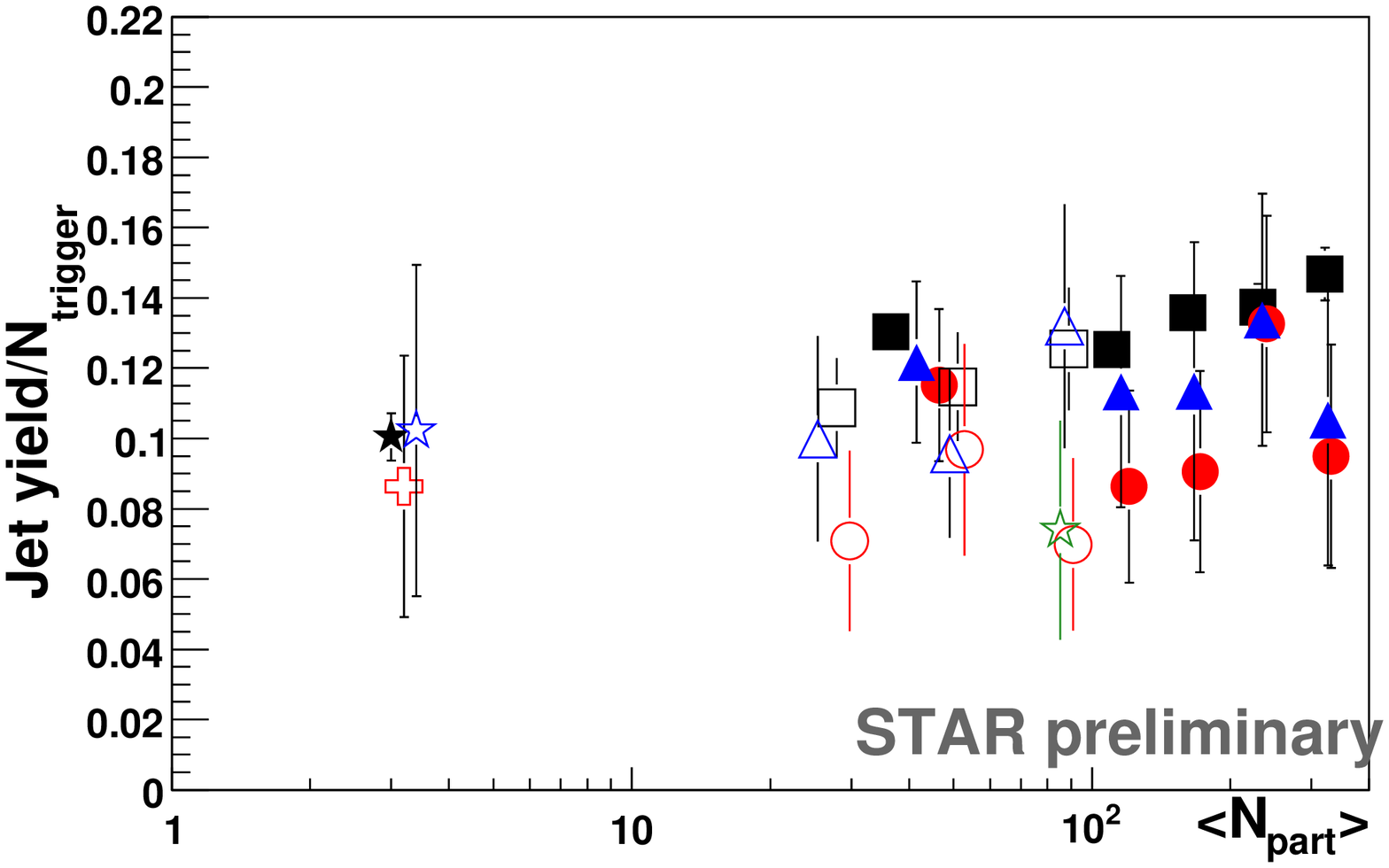}
\end{tabular}
\end{center}
\caption{Centrality and trigger particle type dependence of ridge (left) and jet (right) yields 
in Cu+Cu and Au+Au collisions at $\sqrt{s_{NN}}$~=~200~GeV. The $p_T$ selection criteria used are: 
$p_T^{trig}$~=~3-6~GeV/$c$ and 1.5~GeV/$c$~$<p_T^{assoc}<p_T^{trig}$. The systematic errors 
due to $v_2$ subtraction are shown as a band.} 
\label{jet-ridge-npart}
\end{figure}

\section{Near-side correlation: The ridge}
One of the notable findings in heavy-ion collisions at RHIC has been the observation 
of an additional long-range $\Delta\eta$ correlation on the near-side, {\it the ridge}, which is absent in p+p and d+Au collisions~\cite{MagestroHP,PutschkeHPQM06}. This observation has triggered a lot of interest in theoretical community and a large variety of models explaining the ridge origin have appeared since then~\cite{Chiu:2005ad}.

To quantify the strength of the ridge-like correlations, the measured correlations are decomposed into a jet-like 
component centered at $\Delta\phi\sim$~0, $\Delta\eta\sim$~0 and a ridge component extended in $\Delta\eta$ a top of elliptic flow ($v_2$) modulated background.
It has been observed that the yield of charged particles in the ridge associated with a charged trigger particle increases linearly from peripheral to central Au+Au collisions, while that in jet remains approximately independent of centrality and similar to values in d+Au collisions~\cite{PutschkeHPQM06}. This observation is also confirmed for various trigger particle species studied in Figure~\ref{jet-ridge-npart}~\cite{Bielcikova:2007mb,Nattrass:2008rs}. Within statistical and systematic errors the ridge is independent of the trigger species. Comparing two different collision systems, Cu+Cu and Au+Au, the ridge yield is within errors same at the same number of participants, $N_{part}$~\cite{Nattrass:2008rs}. 

Further, it has been shown that the ridge yield persists up to $p_T\approx$~6~GeV/$c$ and is approximately independent of $p_T^{trig}$~\cite{PutschkeHPQM06}, while the jet yield is increasing with $p_T^{trig}$ in line with jet fragmentation. 
The analysis of the $p_T$ spectra of associated charged particles revealed that the inverse slope extracted from an exponential fit 
is for the ridge-like correlations independent of $p_T^{trig}$ and only by $\approx$~50~MeV 
larger than that of the inclusive $p_T$ spectrum. 

\begin{figure}[t!]
\begin{center}
\begin{tabular}{lll}
\includegraphics[width=4.6cm]{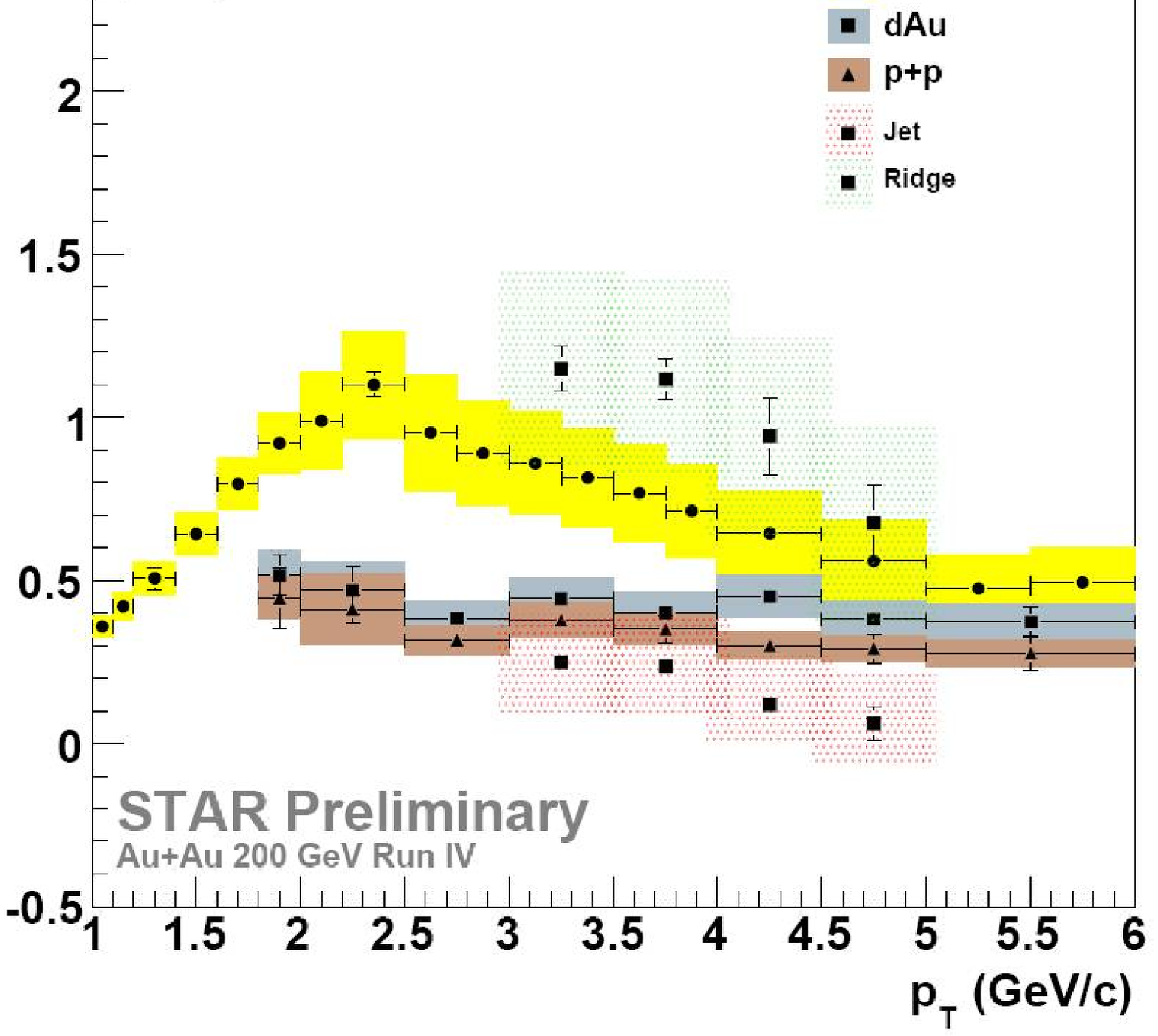}
&
\hspace{0.2cm}
\includegraphics[width=5.7cm]{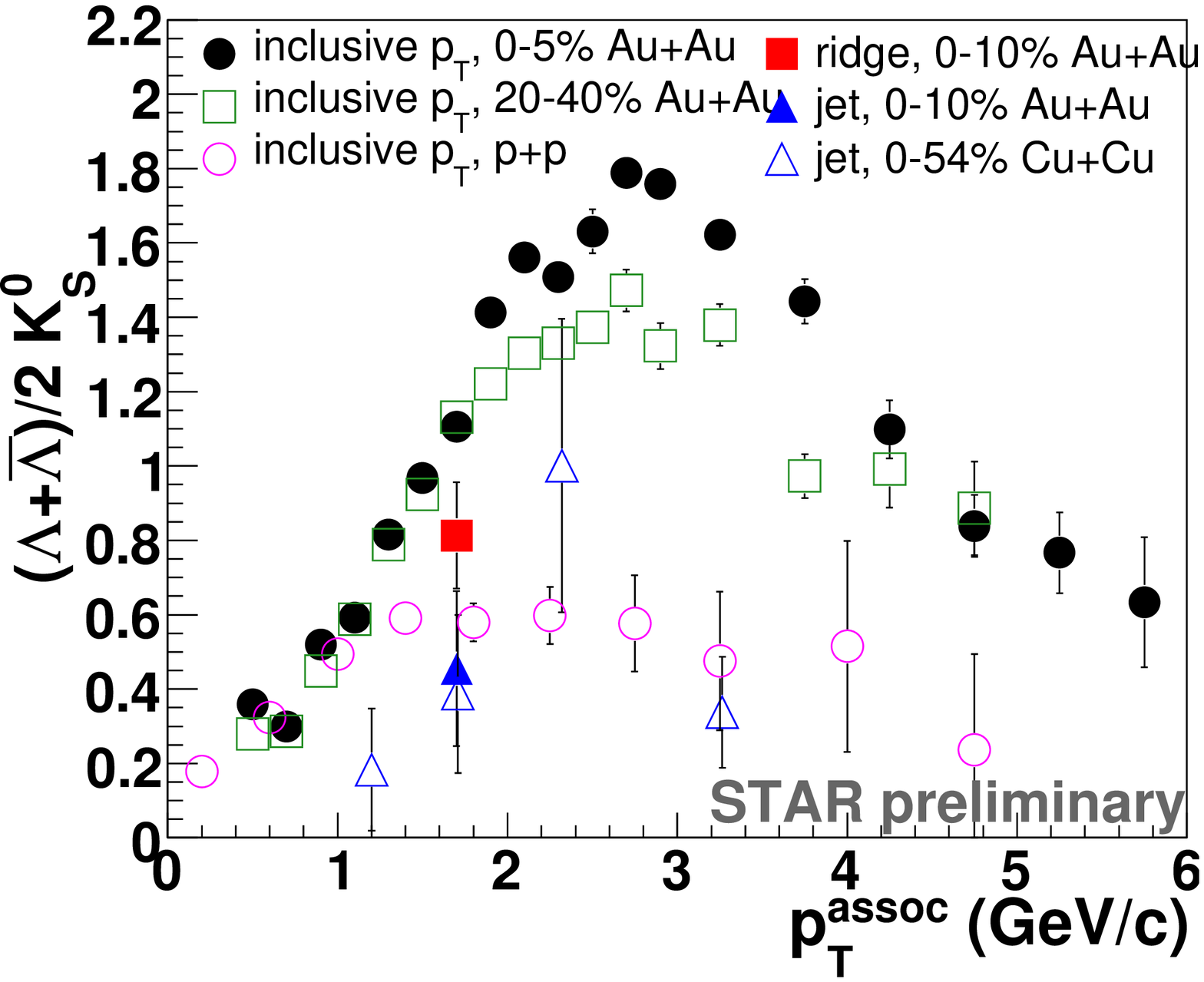}
&
\begin{minipage}[t!]{3.5cm}
\vspace{-5.7cm}
\caption{Baryon/meson ratios in the ridge and jet compared to the inclusive particle ratios at $\sqrt{s_{NN}}$~=~200~GeV: p/$\pi$ (left), $(\Lambda+\bar{\Lambda})/(2K^0_S)$ (right).}
\label{jet-ridge-bmratio}
\end{minipage}
\end{tabular}
\end{center}
\end{figure}

Next the results on particle composition in the jet and ridge are discussed. Figure~\ref{jet-ridge-bmratio} shows the $p_T$ dependence 
of p/$\pi$ and $\Lambda$/$K^0_S$ ratios in the jet and ridge together with the values from inclusive $p_T$ spectra~\cite{Bielcikova:2007mb,Nattrass:2008rs,Suarez}. While the baryon/meson ratios
 for the jet agree with the ratio  measured in p+p, the ratios in the ridge are similar to that from the inclusive measurements. This observation thus supports 
ridge models where hadronization is based on parton recombination.

\begin{figure}[t!]
\begin{tabular}{lr}
\includegraphics[height=4.5cm]{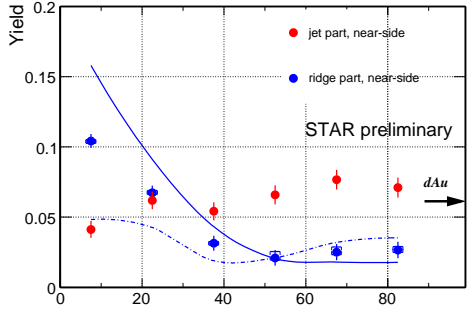}
&
\begin{minipage}[t!]{5.5cm}
\vspace{-5.8cm}
\caption{Ridge and jet yields from di-hadron correlations in non-central (20-60\%) Au+Au collisions at $\sqrt{s_{NN}}$~=~200~GeV as a function of angular difference between the trigger particle and the event plane. The
lines are systematic uncertainties due to elliptic flow subtraction.}
\label{jet-ridge-plane}
\end{minipage}
\end{tabular}
\end{figure}

Figure~\ref{jet-ridge-plane} shows the dependence of the ridge on the orientation of the event plane~\cite{FengQM}. While the ridge yield decreases with the increasing azimuthal angle difference between the trigger particle and the event plane,  
the jet-like yield is constant as the function of the angular difference~\cite{FengQM}. This could indicate a strong near-side 
``jet''-medium interaction in event plane resulting in the ridge formation and a minimal 
interaction perpendicular to the reaction plane.

\begin{figure}[t!]
\vspace{-1.0cm}
\includegraphics[width=17cm]{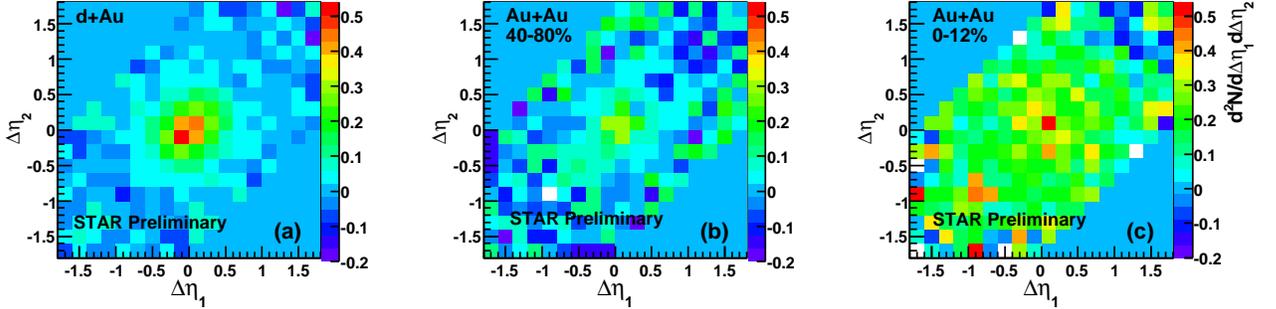}
\caption{Three-particle correlations in $\Delta\eta$ for small azimuthal angle
difference ($\Delta\phi<$~0.7) between associated particles and trigger particle for minimum bias d+Au
(a) and Au+Au (b,c) collisions at $\sqrt{s_{NN}}$~=~200 GeV.}
\label{3part-deta}
\end{figure}

Studies of 3-particle $\Delta\eta_1\times\Delta\eta_2$ correlations, where $\Delta\eta_{1(2)}=\eta^{trig}-\eta^{assoc1(2)}$, are expected to further constrain the physics origin of the ridge. First studies of such correlations for charged particles with $p_T^{trig}$~=~3-10~GeV/$c$ and $p_T^{assoc}$~=1-3~GeV/$c$ are displayed in Figure~\ref{3part-deta}. A clear jet-like peak 
is present  at $(\Delta\eta_1,\Delta\eta_2)\sim$~0 in both, d+Au and Au+Au collisions.  In addition, a uniform overall excess of associated particles is observed in Au+Au collisions implying no correlation (within errors) exists between the particles in the ridge.


\section{The away-side peak: shape modification}

STAR has previously reported a strong suppression of away-side di-hadron correlation peak for $p_T^{trig}$~=~3-6~GeV/$c$ and $p_{T}^{assoc}>$~2~GeV/$c$~\cite{disappearance}. Later it has been observed that this suppression is accompanied by an enhanced production of low-$p_{T}$ associated particles ($p_T^{assoc}>$~0.15~GeV/$c$) resulting in a broader, double peaked distribution at $\Delta\phi\sim\pi$~\cite{reappearance}. Physics mechanisms suggested to explain the away-side peak shape modification include Mach cone shock waves generated by a parton propagating through the medium~\cite{Stoecker:2004qu} or \v{C}erenkov gluon radiation~\cite{Dremin:2005an}. 

\begin{figure}[t!]
\begin{tabular}{lll}
\includegraphics[height=4.5cm]{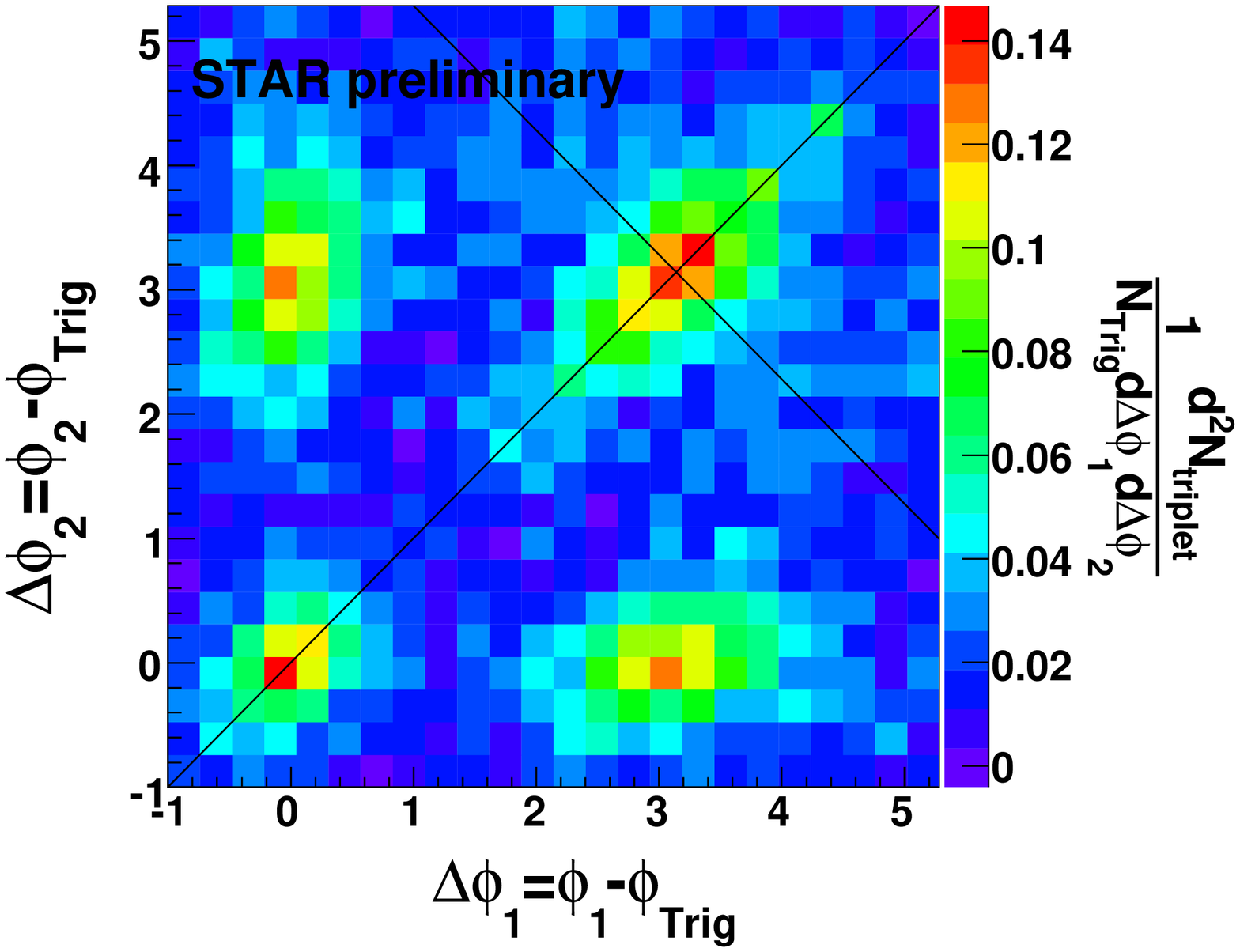}
&
\hspace{-1.3cm}
\includegraphics[height=4.5cm]{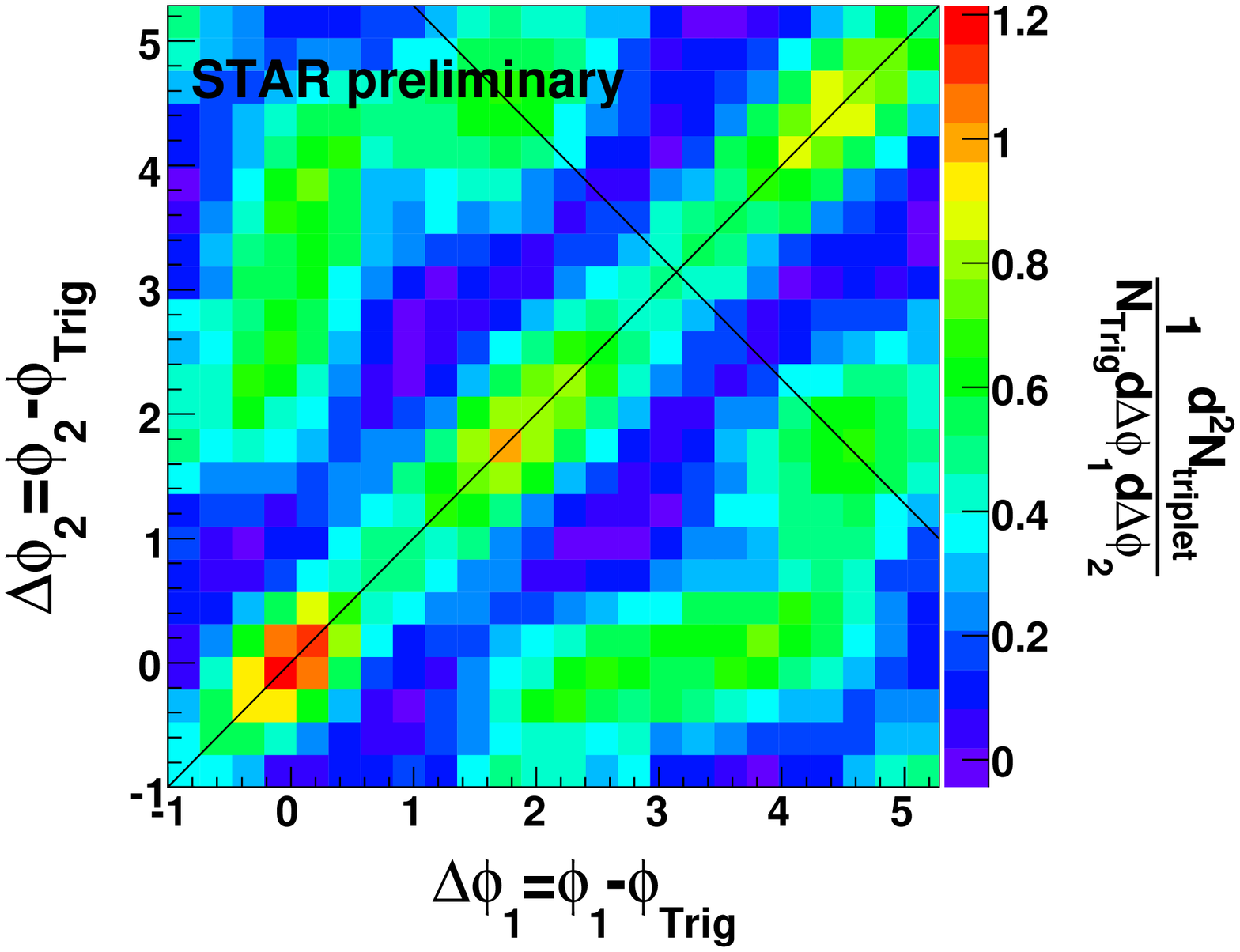}
&
\begin{minipage}[t!]{4.0cm}
\vspace{-5.3cm}
\caption{3-particle azimuthal correlations of charged hadrons for d+Au (left) and 0-12\% 
central Au+Au collisions (right) at $\sqrt{s_{NN}}$~=~200 GeV. 
The trigger particles have $p_T^{trig}$~=~3-4~GeV/$c$ and associated particles have $p_T^{assoc}$~=~1-2~GeV/$c$.}
\label{mach-cone-ulery}
\end{minipage}
\end{tabular}
\end{figure}

\begin{figure}[t!]
\begin{tabular}{lr}
\includegraphics[height=5.0cm]{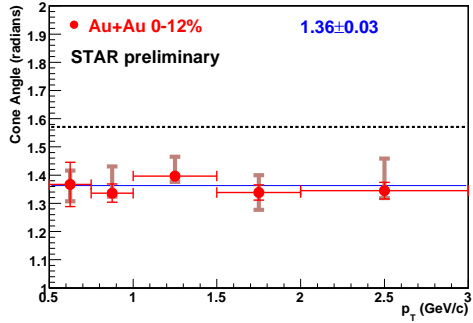}
&
\begin{minipage}[t]{6.5cm}
\vspace{-4.5cm}
\caption{Angles from fits to off-diagonal projections in Figure~\ref{mach-cone-ulery} as a function of $p_{T}^{assoc}$ in 0-12\% central Au+Au collisions at $\sqrt{s_{NN}}$~=~200~GeV for 3~$<p_{T}^{trig}<$~4~GeV/$c$.  The shaded errors are systematic.  The solid line is a fit to a constant giving a cone angle of 1.36$\pm$0.03~radians.}
\label{ulery-projections}
\end{minipage}
\end{tabular}
\end{figure}

In order to distinguish among various physics mechanisms, we investigate 
3-particle $\Delta\phi_{1}\times\Delta\phi_{2}$ azimuthal correlations using one trigger and two associated particles  
The background subtracted correlations in d+Au and central Au+Au collisions 
are shown in Figure~\ref{mach-cone-ulery}~\cite{UleryData,machpaper}. The background has been subtracted using a two component model assuming that one component (``jet'') is correlated 
with the trigger and the other is not except indirect correlations via anisotropic flow~\cite{Ulery06}. 
In Au+Au collisions, the associated particles on the away side populate a cone around $\Delta\phi=\pi$ with 4 
peaks 1.4 radians away from $\Delta\phi=\pi$, which are not observed 
in d+Au collisions. The observed cone angle is independent of collision centrality and $p_{T}^{assoc}$ (cf. Figure~\ref{ulery-projections}), 
favouring the physics picture based on shock waves rather than \v{C}erenkov gluon radiation, in which the cone angle would  decrease with increasing $p_T^{assoc}$.
A model independent analysis based on 3-particle cumulants~\cite{Pruneau06,PruneauData}, has not seen a clear evidence of 
such a conical emission but it is not excluded that the conical emission is masked by flow terms, or too weak to be visible in the cumulant analysis. 
We remark, that the effects of momentum conservation~\cite{Borghini} for the $p_T$ selection used were not accounted for in either of the mentioned analyses.

\section*{Acknowledgments}
The investigations have been partially supported by the IRP AVOZ10480505, by the Grant
Agency of the Czech Republic under Contract No. 202/07/0079 and by the grant LC07048 of
the Ministry of Education of the Czech Republic.

\end{document}